\documentclass{aa}  
\usepackage{graphicx}
\usepackage{natbib}
\usepackage{txfonts}
\usepackage{array}
\usepackage{rotating}
\def\tc{$\theta$\,Car}
\def\xmm{{\sc XMM}\emph{-Newton}}
\def\kms{km\,s$^{-1}$}

\begin{document}
   \title{High-resolution X-ray spectroscopy of \tc\thanks{Based on observations collected with \xmm, an ESA Science Mission with instruments and contributions directly funded by ESA Member States and the USA (NASA).}}

   \author{Ya\"el Naz\'e
          \inst{1}\fnmsep\thanks{Postdoctoral Researcher FNRS}
          \and
          Gregor Rauw\inst{1}\fnmsep\thanks{Research Associate FNRS}
          }

   \offprints{Y. Naz\'e}

   \institute{Institut d'Astrophysique et de G\'eophysique, Universit\'e de Li\`ege, All\'ee du 6 Ao\^ut 17 Bat. B5C, B4000-Li\`ege, Belgium\\
              \email{naze@astro.ulg.ac.be}
             }


  \abstract
   {The peculiar hot star \tc\ in the open cluster IC\,2602 is a blue straggler as well as a single-line binary of short period (2.2d).}  
   {Its high-energy properties are not well known, though X-rays can provide useful constraints on the energetic processes at work in binaries as well as in peculiar, single objects. }
   {We present the analysis of a 50\,ks exposure taken with the \xmm\ observatory. It provides medium as well as high-resolution spectroscopy. }
   {Our high-resolution spectroscopy analysis reveals a very soft spectrum with multiple temperature components (1--6\,MK) and an X-ray flux slightly below the `canonical' value ($\log[L_X(0.1-10.)/L_{BOL}]\sim-7$). The X-ray lines appear surprisingly narrow and unshifted, reminiscent of those of $\beta$\,Cru and $\tau$\,Sco. Their relative intensities confirm the anomalous abundances detected in the optical domain (C strongly depleted, N strongly enriched, O slightly depleted). In addition, the X-ray data favor a slight depletion in neon and iron, but they are less conclusive for the magnesium abundance (solar-like?). While no significant changes occur during the \xmm\ observation, variability in the X-ray domain is detected on the long-term range. The formation radius of the X-ray emission is loosely constrained to $<$5\,R$_{\odot}$, which allows for a range of models (wind-shock, corona, magnetic confinement,...) though not all of them can be reconciled with the softness of the spectrum and the narrowness of the lines.}
   {}

   \keywords{X-rays: stars --  stars: individual: \tc -- stars: early-type}

\titlerunning{X-ray investigation of \tc}
   \maketitle
%

\section{Introduction}

\tc\ (=HD\,93030) is a luminous star of type B0.2V belonging to the open cluster IC\,2602, situated at 152\,pc \citep{rob99}. The cluster is 30\,Myr old, and the massive, hot \tc\ therefore appears to be a rare example of blue straggler. Its optical spectrum display hints of enhanced nitrogen (a three-fold enrichment with respect to solar) and depleted carbon (at least by an order of magnitude) and oxygen (a factor of about 1/4, see \citealt{hub08}). In addition, \tc\ is also a binary system, with short period (2.2d, \citealt{llo95}) and small eccentricity ($e$=0.13, \citealt{hub08}). The companion remains undetected in the visible spectrum (its flux contributes to $<$0.1\% of the total flux) and should therefore be of a much later spectral type ($M\sim1$\,M$_{\odot}$, \citealt{hub08}). The binarity and anomalous abundances might suggest that the blue straggler character of \tc\ results from a past episode of mass-transfer between these two stars.\\

Because of its peculiar properties, the spectrum of \tc\ was investigated several times, notably to search for the presence of a magnetic field \citep{bor79,hub08}. The results are rather inconclusive, but an intriguing period of about 9 minutes was detected in the most recent spectropolarimetric results. \\

We decided to further study the star in the high-energy domain. As it was detected by Einstein and Rosat, \tc\ is known as the brightest X-ray source of IC\,2602, but only approximate X-ray properties have up to now been derived. The current generation of X-ray facilities (Chandra, \xmm) provides a detailed insight on the X-ray emission, especially thanks to their grating instruments. However, these observatories have observed only a handful of B stars at high spectral resolution ($\tau$\,Sco, $\beta$\,Cru, $\epsilon$\,Ori, Spica) and no blue straggler. The analysis of \tc\ thus nicely fills a gap, and could help better understand the high-energy characteristics of hot stars.\\

The paper is organized as follows: the data and reduction processes are presented in Sect. 2, the general properties of \tc\ in the X-ray domain are examined in Sect. 3.1, the detailed characteristics of the X-ray lines are derived in Sect. 3.2, and we finally conclude in Sect. 4.

\section{X-ray Observations}

On 2002 Aug. 13 (Rev. 0490, PI R.Pallavicini), the IC\,2602 cluster was observed with \xmm\ for a total exposure time of 50\,ks. We retrieved this dataset from the \xmm\ public archives, in order to perform a thorough analysis of the high-energy emission of \tc. We processed these archival data with the Science Analysis System (SAS) software, version~7.0; further analysis was performed using the FTOOLS tasks and the XSPEC software v 11.2.0. \\

For this observation, the three European Photon Imaging Cameras (EPICs) were operated in the standard, full-frame mode and a thick filter was used to reject optical light. After application of the pipeline chains (tasks {\sc emproc, epproc}), the EPIC data were filtered as recommended by the SAS team: for the EPIC MOS (Metal Oxide Semi-conductor) detectors, we kept single, double, triple and quadruple events (i.e.\ pattern between 0 and 12) that pass through the \#XMMEA\_EM filter; for the EPIC pn detector, only single and double events (i.e.\ pattern between 0 and 4) with flag$=$0 were considered. To check for contamination by low-energy protons, we have further examined the light curve at high energies (Pulse Invariant channel number$>$10000, E$\gtrsim$10\,keV, and with pattern$=$0). A large background flare due to soft protons (most probably of solar origin) occured towards the end of the observation and we therefore discarded time intervals where the lightcurves for PI$>$10\,000 present an EPIC MOS count rate larger than 0.3 cts\,s$^{-1}$ and an EPIC pn count rate larger than 1 cts\,s$^{-1}$. The pn data of \tc\ are severely affected by the presence of CCD gaps and bad columns and we therefore decided to discard them from the present analysis. A circular source region of 50'' radius and an annular background region of 50 and 90'' radii (both centered on \tc) were then used to extract spectra and lightcurves. Due to the brightness of the source, it was not necessary to bin the EPIC MOS spectra, but bad energy bins and noisy bins at E$<$0.15\,keV or $>$10.\,keV were discarded. \\

In this observation, \tc\ appears near the center of the field-of-view (FOV) and some information could therefore be recorded with the Reflection Grating Spectrograph (RGS). However, since \tc\ was not {\it exactly} at the center of the FOV, the standard processing could not be used: the pipeline chain (task {\sc rgsproc}) had to be run specifically with the coordinates of the object. As for EPIC, a flare in the lightcurve was detected and time intervals presenting an RGS count rate larger than 0.35 cts\,s$^{-1}$ were therefore discarded. In addition, a relatively bright X-ray source associated with HD\,307938 (G0) also appears in the RGS FOV: though it does not contaminate the RGS spectrum of \tc, it could lead to a wrong estimate of the background and we therefore had to mask this source for the background calculation. For both non-standard operations, we followed closely the recommendations from the SAS team (see the SAS threads on the XMM website). The RGS spectra were binned to reach at least 5 cts per bin and only data between 0.35 and 1.8\,keV were considered; the data from the second order contain too few counts to be useful, and we therefore also discarded them from the analysis.

\section{High-energy properties of \tc}

\subsection{General characteristics}

\begin{figure}
\begin{center}
\includegraphics[width=8cm]{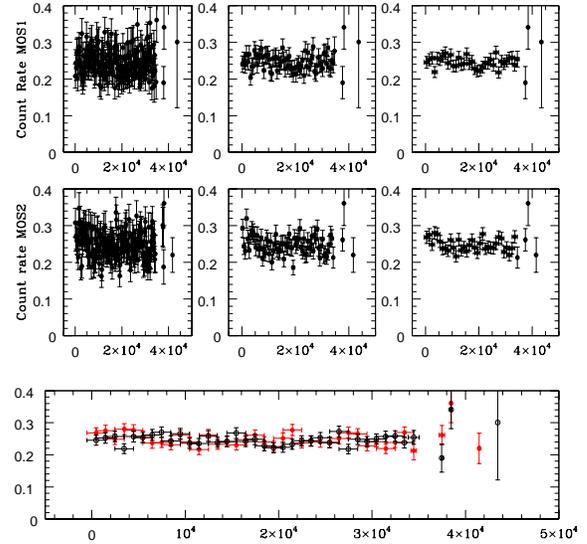}
\caption{\label{tc_lc} X-ray lightcurves of \tc\ in the 0.2--5\,keV band. {\it Top:} MOS1 lightcurves with time bins of 200, 500 and 1000s (from left to right). {\it Middle:} Same for MOS2. {\it Bottom:} Superposition of the MOS1 (open black symbols) and MOS2 (filled red symbols) lightcurves with 1ks-bins.} 
\end{center}
\end{figure}

\tc\ appears as the brightest X-ray source in the field of IC\,2602. The SAS detection algorithm (task {\sc edetect\_chain}) indicates that the count rates in the 0.4--10\,keV energy band amount to 0.255$\pm$0.003 cts\,s$^{-1}$ for MOS1 and to 0.258$\pm$0.003 cts\,s$^{-1}$ for MOS2 (note that this is still well below the limit for pile-up). In addition, the hardness ratios are very small, revealing the softness of the source:  HR1=(M-S)/(M+S) is $-$0.720$\pm$0.008 for MOS1 and $-$0.724$\pm$0.008 for MOS2, HR2=(H-M)/(H+M) is $-$0.920$\pm$0.015 for MOS1 and $-$0.956$\pm$0.011 for MOS2 where S, M, and H are the count rates in the soft (0.4--1\,keV), medium (1--2\,keV) and hard (2--10\,keV) energy bands. Finally, the MOS lightcurves in the 0.2--5\,keV band were analyzed by $\chi^2$ and $pov$ tests \citep{san04} but no significant variation was detected during the 50 ks observation (Fig. \ref{tc_lc}, note that this covers about one quarter of the orbit of the spectroscopic binary). We thus decided to analyze the spectral properties derived from the whole observation. 

\begin{figure*}
\begin{center}
\includegraphics[width=8cm,height=7cm]{thetacar_epic.ps}
\includegraphics[width=9cm]{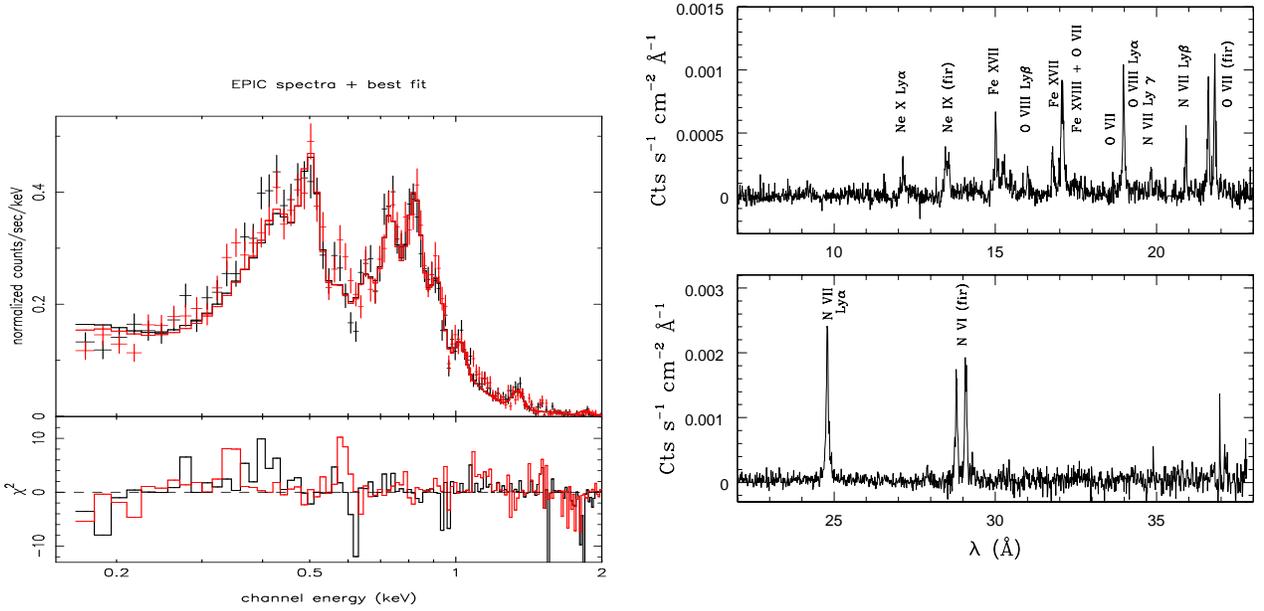}
\caption{\label{tc_spec} {\it Left:} EPIC MOS spectra of \tc\ with the best-fit RGS+EPIC model superimposed (see Table \ref{spec} - MOS1 appearing in black, MOS2 in red). {\it Right:} Fluxed RGS spectrum of \tc\ (from task {\sc rgsfluxer}, RGS1+2, orders 1+2) with the main lines identified. Note that the scales of the y-axis differ by a factor of 2 between the upper and lower panels.} 
\end{center}
\end{figure*}

Two types of spectra are available: the EPIC low-resolution ones and the RGS high-resolution ones (Fig. \ref{tc_spec}). Both datasets show that the majority of the flux is clearly confined to 0.1--2\,keV, confirming the softness of the source. \\

The thermal nature of the source is further revealed by the presence of lines, without any significant continuum, in the RGS spectra. These lines are triplets from He-like ions (Ne\,{\sc ix}, N\,{\sc vi}, and O\,{\sc vii}), Lyman lines from H-like ions (Ne\,{\sc x}, N\,{\sc vii}, and O\,{\sc viii}) and features associated with ionized iron. Their emissivities peak at temperatures of 1.4--5.8MK (0.12--0.50\,keV). A striking characteristics of the spectrum of \tc\ is readily derived from a visual comparison of the line strengths with those of `normal' OB stars (see e.g. $\zeta$\,Ori in Fig. 1 of \citealt{pol07}): the nitrogen lines appear much stronger while the oxygen and neon lines are weaker. In this context, it should be noted that the carbon line at 33.7\AA\ and the magnesium line at 9.2 \AA\ are below our detection level. Finally, the X-ray lines appear relatively narrow: FWHM$\sim$1000\,\kms\ for O\,{\sc viii}\,Ly$\alpha$\,$\lambda$\,18.97\AA\ and N\,{\sc vii}\,Ly$\alpha$\,$\lambda$\,24.78\AA, $\sim$700\,\kms for N\,{\sc vii}\,Ly$\beta$\,$\lambda$\,20.91\AA, i.e. the observed line width is mainly instrumental and does not reflect the intrinsic line width (see also results of the fits below). \\

   \begin{table*}
      \caption{Best-fit models and X-ray fluxes at Earth for \tc. For each parameter, the lower and upper limits of the 90\% confidence interval (derived from the {\sc error} command under {\sc xspec}) are noted as indices and exponents, respectively. The normalisation factors are defined as $10^{-14}\int n_e n_{\rm H} dV/4\pi D^2$, where $D$, $n_e$ and $n_{\rm H}$ are respectively the distance to the source, the electron and proton density of the emitting plasma; the abundance parameters are ratios of the abundance by number (with respect to hydrogen) between the observed spectrum and the solar value. The models are of the type $wabs\times vapec$ or $wabs\times(vapec+vapec)$, with $N^{\rm H}$ fixed to 1.8$10^{20}$~cm$^{-2}$ \citep{dip94}. The observed fluxes are given in the 0.5--10\,keV band and in units $10^{-12}$~erg\,cm$^{-2}$\,s$^{-1}$.}
         \label{spec}
     \centering
         \begin{tabular}{lcccccccccccc}
            \hline\hline
Instr. & k$T_1$ & norm$_1$ & k$T_2$ & norm$_2$ & C & N & O & Ne & Mg & Fe & $\chi^2_{\nu}$ (dof) & $f_{\rm X}^{\rm abs}$ \\
& keV   & $10^{-4}$cm$^{-5}$ & keV   & $10^{-4}$cm$^{-5}$        & & & & & &  \\
            \hline
\vspace*{-0.3cm}&\\
MOS1+2 & 0.15$_{0.14}^{0.16}$ & 9.10$_{7.12}^{10.8}$ & 0.36$_{0.35}^{0.38}$ & 7.30$_{5.56}^{9.35}$ & 2.22$_{1.56}^{3.37}$ & 5.04$_{4.17}^{6.43}$ & 0.36$_{0.29}^{0.47}$ & 1.19$_{0.95}^{1.52}$ & 1.23$_{0.95}^{1.62}$ & 0.96$_{0.78}^{1.22}$ & 0.40  (1300) & 1.48 \\
\vspace*{-0.3cm}&\\
RGS 1+2& 0.24$_{0.24}^{0.25}$ & 28.4$_{23.6}^{32.8}$ &  &  & 0.$_{0.}^{0.04}$ & 2.32$_{1.93}^{2.90}$ & 0.10$_{0.08}^{0.12}$ & 0.38$_{0.31}^{0.50}$ & 2.47$_{1.19}^{4.07}$ & 0.61$_{0.50}^{0.78}$ & 1.85 (594) & 1.20\\
\vspace*{-0.3cm}&\\
RGS 1+2& 0.16$_{0.15}^{0.22}$ & 6.84$_{4.92}^{23.2}$ & 0.39$_{0.37}^{0.41}$ & 16.0$_{5.97}^{19.2}$ & 0.$_{0.}^{0.09}$ & 3.85$_{2.98}^{4.99}$ & 0.19$_{0.10}^{0.25}$ & 0.40$_{0.30}^{0.51}$ & 0.83$_{0.30}^{1.43}$ & 0.32$_{0.26}^{0.40}$ & 1.74 (592) & 1.35\\
\vspace*{-0.3cm}&\\
RGS+MOS& 0.16$_{0.15}^{0.16}$ & 12.8$_{10.9}^{14.8}$ & 0.32$_{0.31}^{0.32}$ & 23.1$_{20.2}^{26.3}$ & 0.05$_{0.}^{0.12}$ & 2.32$_{2.13}^{2.54}$ & 0.12$_{0.11}^{0.13}$ & 0.40$_{0.36}^{0.45}$ & 0.51$_{0.42}^{0.60}$ & 0.38$_{0.34}^{0.41}$ & 0.93 (1902) & 1.44\\
\vspace*{-0.3cm}&\\
            \hline
         \end{tabular}
   \end{table*}

   \begin{table}
      \caption{Abundance constraints from optical spectra (adapted from \citealt{hub08}). The helium abundance is given as the ratio in numbers with respect to hydrogen whereas the others are given in logarithmic value (hydrogen abundance conventionally being at 12.0 in this case); {\sc xspec} solar abundances are those of \citet{and89}.}
         \label{abond}
     \centering
         \begin{tabular}{lccc}
            \hline\hline
Element & El. Abond. & El. Abond. & ratio  \\
        & in vis. & in {\sc xspec} & \\
            \hline
He   & 0.083$\pm$0.028 & 0.0977 & $\sim$1\\
C    & 7.16$\pm$0.46   & 8.56   & 0.04\\
N    & 8.56$\pm$0.27   & 8.05   & 3.24\\
O    & 8.38$\pm$0.22   & 8.93   & 0.28\\
Si   & 7.43$\pm$0.23   & 7.55   & $\sim$1\\
S    & 7.32$\pm$0.33   & 7.21   & $\sim$1\\
            \hline
         \end{tabular}
   \end{table}

The EPIC MOS spectra present two broad maxima, at about 0.45\,keV and 0.8\,keV. While one might then expect the presence of two dominant temperatures, these two peaks actually correspond to the two main groups of lines as revealed by RGS data (strong N lines at long wavelengths for the lower-energy peak and numerous ONeFe lines for the other one). A fit to these data with a differential emission measure model (DEM, $c6pvmkl$, \citealt{lem,sin}) yields only one single, broad thermal component with a peak at 0.2\,keV, and an interval of the temperature distribution at half maximum between 0.15 and 0.5\,keV, in nice agreement with the peak emissivity temperatures (see above). \\

Unfortunately, the $mekal$ model can not reproduce well the exact rest wavelengths of the X-ray lines, as is obvious from a comparison with the high-resolution data, and we must therefore rely on the more recent and more precise $apec$ model. Since no DEM model of the $apec$-type is available in {\sc xspec}, we tried to fit the spectra with single temperature components. For EPIC MOS data, an absorbed 1T fit does not yield $\chi^2<2$ but absorbed 2T models give good results (absorbed 3T models do not yield any significant improvement of the fits). Since lines of NONeFe are clearly detected while conspicuous lines of C and Mg are absent, we let the abundances of these 6 elements free to vary. We fixed the other abundances to the solar value since no strong constraining information (presence/absence of lines) can be derived from the data. In addition, the absorbing column was fixed to the interstellar value derived in the UV (1.8$10^{20}$~cm$^{-2}$ \citealt{dip94}) since the free-$N^{\rm H}$ fits do not suggest any significant additional, circumstellar absorption. For RGS data, both 1T and 2T yield acceptable residuals. In addition to varying the usual $apec$ parameters, two trials were attempted. First, varying the redshift parameter in order to reproduce a global velocity shift: the best fit was found for positive velocities of +60--100\,\kms, values which are insignificant given the RGS resolution (see also below). Second, reproducing a significant intrinsic width of the lines by convolving the $apec$ model by a gaussian ($gsmooth$ model): no significant intrinsic width could be found, reinforcing our conclusion of line widths being mainly instrumental in nature for \tc. \\

The parameters of the best-fit models are listed in Table \ref{spec}. To interpret these results, some facts must be kept in mind: (1) the RGS data are the most sensitive to abundance variations since the lines are resolved for these spectra whereas the line blends observed in EPIC may yield unrealistic values in some cases (see e.g. the case of carbon\footnote{Indeed, a large overabundance in carbon, as derived from the EPIC MOS fits alone, would result in the presence of a strong carbon line at 35\AA, which is not observed in the RGS data.}), (2) the EPIC data are the best to constrain the absorbing column and high-temperature components. Overall, the results of the fits are twofold. On the one hand, the fitted temperatures are low (0.15+0.4\,keV or 0.24\,keV). No hard tail is present, nor any high temperature: this rules out any significant contamination by a putative close PMS star (like in the case of $\beta$\,Cru\,D, \citealt{coh08}) or from the low-mass companion \citep[for a review of the X-ray emission of solar-like stars, see][]{gud07}. On the other hand, abundances are clearly non-solar: N enhanced by at least a factor of 2, C at only a few percent of the solar values, ONeFe also depleted (though to a lesser extent than carbon) - the case of magnesium unfortunately appears unclear due to rather erratic results. This abundance pattern measured on the RGS data is close to the results derived from the visible spectrum (see Table \ref{abond}). \\

The unabsorbed flux (i.e. the flux corrected for the interstellar absorption) measured on the RGS+MOS model amounts to 3.8$\times10^{-12}$~erg\,cm$^{-2}$\,s$^{-1}$ in the 0.1--10\,keV energy band or 1.6$\times10^{-12}$~erg\,cm$^{-2}$\,s$^{-1}$ in the 0.5--10\,keV energy band, resulting in luminosities of 1.1$\times10^{31}$~erg\,s$^{-1}$ and 4.4$\times10^{30}$~erg\,s$^{-1}$, respectively, for a distance of 152pc \citep{rob99}. For the 0.5--10\,keV energy range, this converts into a $\log(L_X/L_{BOL})$ of $-$7.35\footnote{A bolometric luminosity of 9.7$\times10^{37}$~erg\,s$^{-1}$ was derived from B=2.54, V=2.78, E(B-V)$\sim$0., a distance of 152pc, and a bolometric correction of $-$3.16. It is compatible with the values found in the literature ($\log[L/L_{\odot}]$=4.30--4.54, \citealt{gat81,var85,prin89}).}, or $-$6.97 taking into account the 0.1--0.5\,keV flux. These ratios are slightly lower than the `canonical' value measured for O and early B stars \citep[$-$6.9 for the 0.5--10.\,keV band, ][]{san06}: \tc\ is thus not a bright X-ray emitter for its class, unlike what has been sometimes claimed. \\

\tc\ was observed by both Einstein and ROSAT. It appears in the 2E catalog \citep{har94} with an IPC count rate of 0.089$\pm$0.006 cts s$^{-1}$ and in the RASS bright source catalog \citep{vog99} with a PSPC count rate of 0.42$\pm$0.04 cts s$^{-1}$. For ROSAT, \citet{cas94} further reported a count rate of 0.374$\pm$0.016 cts s$^{-1}$, a value compatible with the RASS detection\footnote{\citet{rand95} quote 0.299$\pm$0.002 cts s$^{-1}$ for raster scans but the authors mentioned the discrepancy and suggest the scans may yield more uncertain values}. By folding the best fit RGS+EPIC model through the response matrices of these past facilities, we can estimate the ``equivalent" IPC or PSPC count rate of this \xmm\ observation: 0.119 cts s$^{-1}$ for IPC and 0.326 cts s$^{-1}$ for PSPC. During the \xmm\ observation, \tc\ was thus 30\% times brighter than during the IPC exposure (a 5$\sigma$ variation) and $\sim$20\% times fainter than during the PSPC exposures (a 3$\sigma$ variation). Though \tc\ does not present any short-term variability during the 50\,ks exposure, some longer-term changes are apparently present and future observations are needed to ascertain their nature: are they truly long-term variations, unrelated to the orbital period, or do they occur during the unobserved part of the orbit? \\

\subsection{Analysis of the X-ray lines}

Since there are a few strong, isolated lines recorded on the RGS spectra, we decided to fit these individually using gaussians (including a constant for the pseudo-continuum) within {\sc xspec} and {\sc midas}. The results of these fits can be found in Table \ref{lines}. First, we focused on single lines of the Lyman series. Shifts detected by individual fits are small - in fact below the detection limit considering the RGS characteristics. They broadly agree with the free-redshift fits (see above), small differences being expected since the Aped and Spex databases are not identical. In addition, no significant asymetry, or intrinsic broadening ($\sigma_{int}<$350\,\kms) were detected. Three other strong X-ray lines actually are $fir$ triplets of He-like ions, i.e. they are composed of a forbidden line, two close intercombination lines and a resonance line. It should be noted that thermal plasma models like $apec$ were only able to reproduce the {\it r} component of these triplets. A first fit revealed no significant shift/width variations between the lines, as expected theoretically (the lines being formed under the same physical conditions). Therefore, the intrinsic widths and velocities of the three components were constrained to be equal (as for single lines, they were subsequently found to be undetectable).  Narrow, unshifted lines are not common amongst hot stars and are often considered as unreconciliable with a wind-shock origin for the X-ray emission. Such lines are generally expected for magnetic objects ($\theta^1$\,Ori\,C, $\tau$\, Sco, \citealt{gag05,mew}), a notable execption being another early B-type star, $\beta$\,Cru that has narrow lines but no magnetic field \citep{coh08}. \\ 

   \begin{table*}
      \caption{Best-fit gaussians for selected X-ray lines. Rest energies of the lines are taken from the Spex database v2.0 (Kaastra, 2002, private communication). Ly$\alpha$ lines contain two components too close to be disentangled; the velocities given below are calculated with respect to the A component. Emission measures are calculated using emissivities at the mean plasma temperature ($kT$=0.2\,keV, or $\log T$=6.4) or the maximum emissivities - both being taken from the Spex database.}
         \label{lines}
     \centering
         \begin{tabular}{lcccccc}
            \hline\hline
Ion & $E_0$ & $v$ & $\sigma$ & $f_{\rm X}^{\rm abs}$ & EM ($kT$=0.2\,keV) & EM (max emissivity)\\
& keV & \kms & $10^{-4}$keV   & $10^{-5}$\,phot\,cm$^{-2}$\,s$^{-1}$  & $10^{53}$\,cm$^{-3}$ & $10^{53}$\,cm$^{-3}$\\
            \hline
\multicolumn{7}{l}{\it H-like lines}\\
\vspace*{-0.3cm}&\\
O\,{\sc viii}\,Ly$\alpha$ & 0.653550 & $-112_{-129}^{-74}$  & $0._{0.}^{2.61}$     & 8.41$_{7.20}^{9.63}$ & 1.05$_{0.90}^{1.20}$ & 0.89$_{0.76}^{1.01}$\\
\vspace*{-0.3cm}&\\
N\,{\sc vii}\,Ly$\beta$   & 0.592940 & $-24_{-324}^{233}$   & $0._{0.}^{6.85}$     & $2.73_{1.61}^{4.00}$ & 25.3$_{14.9}^{37.1}$ & 24.2$_{14.3}^{35.4}$\\
\vspace*{-0.3cm}&\\
N\,{\sc vii}\,Ly$\alpha$  & 0.500360 & $-218_{-223}^{-108}$ & $0._{0.}^{1.64}$     & 24.5$_{22.5}^{26.4}$ & 21.2$_{19.5}^{22.9}$ & 18.6$_{17.1}^{20.1}$\\
\vspace*{-0.3cm}&\\
\multicolumn{7}{l}{\it He-like lines}\\
\vspace*{-0.3cm}&\\
Ne\,{\sc ix} f& 0.90499& $-374_{-382}^{+162}$  & $0._{0.}^{6.53}$   & $0._{0.}^{0.24}$     & 4.71$_{3.43}^{7.44}$ & 2.25$_{1.64}^{3.55}$\\
\vspace*{-0.3cm}&\\
Ne\,{\sc ix} i& 0.91481&                      &                    & $1.92_{1.25}^{3.40}$ \\
\vspace*{-0.3cm}&\\
Ne\,{\sc ix} r& 0.92195&                      &                    & $2.92_{2.27}^{4.00}$ \\
\vspace*{-0.1cm}&\\
O\,{\sc vii} f& 0.56101& $4_{-112}^{+48}$      & $0._{0.}^{1.68}$   & $0.67_{0.}^{1.57}$   & 1.25$_{0.95}^{1.59}$ & 1.08$_{0.81}^{1.37}$\\
\vspace*{-0.3cm}&\\
O\,{\sc vii} i& 0.56874&                      &                    & $8.38_{6.65}^{10.0}$ \\
\vspace*{-0.3cm}&\\
O\,{\sc vii} r& 0.57395&                      &                    & $6.73_{5.27}^{8.50}$ \\
\vspace*{-0.1cm}&\\
N\,{\sc vi} f & 0.41986& $-34_{-57}^{+43}$     & 0.$_{0.}^{0.82}$   & $0._{0.}^{1.01}$     & 63.3$_{55.1}^{73.6}$ & 18.0$_{15.7}^{20.9}$\\
\vspace*{-0.3cm}&\\
N\,{\sc vi} i & 0.42621&                      &                    & $17.1_{15.1}^{19.3}$ \\
\vspace*{-0.3cm}&\\
N\,{\sc vi} r & 0.43065&                      &                    & $14.4_{12.3}^{16.3}$ \\
\vspace*{-0.3cm}&\\
            \hline
         \end{tabular}
   \end{table*}

The relative strengths of the {\it fir} lines depend on the plasma temperature, the plasma density, and the stellar radiation field. More precisely, the {\it f} line is weakened and the {\it i} line strenghtened when the density or the radiation field are high; for hot stars such as \tc, the latter has a much stronger impact than the former. On the other hand, the {\it (f+i)/r} ratio is sensitive to plasma temperature \citep[for more details see e.g.][]{por01}. These {\it fir} lines can not be reproduced by `simple' {\sc xspec} models, as already mentioned, but the gaussian fits allow us to derive the {\it f/i} and {\it (f+i)/r} ratios. Monte-Carlo simulations, assuming a normal distribution for the line fluxes (90\% confidence interval is $\pm$1.645$\sigma$), were used to estimate the errors on these ratios.\\

\begin{figure}
\begin{center}
\includegraphics[width=9cm]{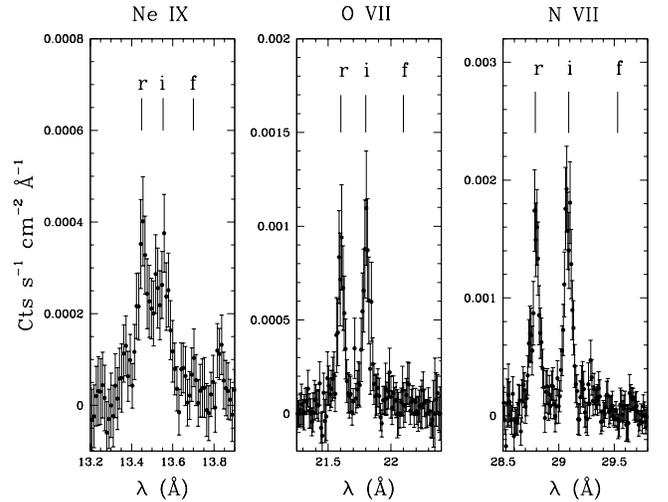}
\caption{\label{tc_fir} Zoom on the {\it fir} triplets recorded in the fluxed RGS spectrum of \tc. Tickmarks indicate the rest wavelengths of the lines. The line at 13.82\AA\ belongs to Fe\,{\sc xvii}; there is no trace of the Fe\,{\sc xix} line at 13.78\AA\ that could contaminate the f component of the Ne\,{\sc ix} triplet.}
\end{center}
\end{figure}

In contrast with the colliding-wind systems where $f\gg i$ \citep{pol05}, the forbidden lines of \tc\ remain undetected in all three triplets (Fig. \ref{tc_fir}). Therefore, only an upper limit can be derived on the {\it f/i} ratios: the one-sided 90\% confidence interval is 0--0.06 for N\,{\sc vi}, 0--0.17 for O\,{\sc vii}  0--0.15 for Ne\,{\sc ix}. These ratios can then be compared with the values from \citet{por01} for $T_e$=2.0MK (close to the main component at k$T\sim$0.2\,keV) $T_{rad}$=30kK (since $T_{eff}$=31$\pm$1kK, \citealt{hub08}) and dilution factors $W$ of 0.01, 0.10 and 0.50: the oxygen and nitrogen ratios do not provide strong constraints (even $W$ values lower than 0.01 are permitted), but the neon triplet favors dilution factors $W$ of $\sim$0.5, i.e. an emission occuring close to the star. Since \tc\ and $\tau$\,Sco display similar stellar parameters (same $T_{eff}$, gravity, spectral type), we can infer more precise limits by using the {\it f/i} evolution calculated by \citet{mew} and shown in Fig. 2 of that paper: the formation radii are $<$20\,R$_*$ from the N\,{\sc vi} and O\,{\sc vii} triplets, but $<$5\,R$_*$ for the Ne\,{\sc ix} triplet. Such values agree well with those found in other hot stars, but they are too loosely constrained to enable us to distinguish between a wind-shock, coronal or magnetic origin for the X-ray emission.\\

Turning to the {\it (f+i)/r} ratio, the values are 1.19$_{1.03}^{1.49}$ for N\,{\sc vi}, 1.34$_{1.01}^{1.90}$ for O\,{\sc vii} and 0.66$_{0.33}^{1.23}$ for Ne\,{\sc ix}, where indices and exponents give the lower and upper limits of the 90\% confidence interval. These values are close to one, suggesting the plasma to be collision-dominated \citep{por01}. Comparison with the theoretical values from \citet{por01} yields plasma temperatures $\lesssim$1\,MK for N\,{\sc vi} and O\,{\sc vii} and $\sim$6\,MK for Ne\,{\sc ix}. This indicates that a range of temperatures could exist in the X-ray emitting plasma, as was already suggested by the rather broad peak of the DEM models. \\

Another independent temperature estimate relies on the ratio of He-like to H-like line fluxes. The lines fluxes were first dereddenned before estimating this ratio. For nitrogen, it amounts to 1.41$\pm$0.14 if one considers the Lyman\,$\alpha$ line or 12.6$\pm$3.6 for Lyman\,$\beta$. Interpolating linearly the Spex database, such ratios correspond to $\log(T)$ of 6.23$\pm$0.02 and 6.25$\pm$0.03, respectively. For oxygen, the He-to-H line ratio amounts to 1.99$\pm$0.36, which yields $\log(T)$=6.37$\pm$0.02. These temperatures agree well with the results from DEM and $apec$ models (see above); they are larger than those derived from the comparison of {\it (f+i)/r} ratios and Porquet et al.'s models but discrepancies between these two temperature estimates are often found. \\

Finally, the observed line fluxes in units ph\,cm$^{-2}$\,s$^{-1}$ were dereddenned and converted to erg\,cm$^{-2}$\,s$^{-1}$, in order to derive the emission measures (EM=4$\pi d^2 f^{unabs}/Q(T_e)$) at a distance of 152pc. The emissivities $Q(T_e)$ were found in the Spex database, and were summed if several components (like the {\it fir} lines) exist. Table \ref{lines} present the emission measures (EMs) calculated at the mean plasma temperature ($\sim$0.2\,keV) and at the temperature of maximum emissivity. It should be noted that O\,{\sc vii} and O\,{\sc viii} lines yield similar EMs, but the value for N\,{\sc vi} is three times that of N\,{\sc vii} if one considers the mean plasma temperature. However, the discrepancy disappears if one considers peak emissivity temperatures. This might be interpreted once again as a need to drop the too simplistic assumption of having only one plasma temperature: most probably, the X-ray emitting plasma displays a range in temperatures, individual lines being emitted at the best conditions for each one. We will thus only consider the EM values calculated with the maximum emissivities. For solar abundances, all EMs should have similar values; the dispersion of the observed EMs confirms once again the non-solar abundances of \tc. When dividing the EMs, we find a N/O enhancement of a factor of 20 and a Ne/O ratio of about 2.5, compared to solar values. The global fits, though they could not reproduce the intense {\it i} component of the triplets, confirm these values. The (N/O)$_*$/(N/O)$_{\odot}$ ratio is only slightly larger than the value of $\sim$12 found by \citealt{hub08}). This underlines the good capabilities of the X-ray data for constraining abundances.

\section{Conclusions}

\xmm\ observations revealed the atypical character of \tc. Overall, its X-ray emission appears very soft as well as rather weak. Almost all the flux is found below 1\,keV; indeed, spectral fits indicate a dominant temperature of about 0.2\,keV. The total unabsorbed luminosity in the 0.1--10\,keV range amounts to 1.0$\times10^{31}$~erg\,s$^{-1}$, which yields a $L_X/L_{BOL}$ ratio slightly lower than the `canonical' value for OB stars. Though no significant variability is detected during the 50\,ks \xmm\ observation, an increase (resp. decrease) of the flux is clearly observed when comparing with previous Einstein (resp. ROSAT) observations. \\

The high-resolution spectrum of \tc, revealed by the RGS, does not show any significant continuum emission but is solely composed of lines of H- and He-like ions of N, O, and Ne as well as some iron lines. To the resolution limits of the RGS, these lines appear narrow and unshifted: the observed width of these lines is mainly instrumental and the intrinsic width is limited to $<350$\,\kms. It must be noted that the nitrogen lines are anomalously strong: the spectral fits reveal a large enrichment in nitrogen (abundance 3 times solar), together with a strong deficit in carbon, which both agree well with the values found in the optical spectrum \citep{hub08}. Neon, oxygen and iron also appear slightly depleted in the X-ray spectrum. For the {\it fir} triplets, the forbidden component is fully suppressed while the intercombination lines are slightly stronger than the resonance component. The {\it f/i} ratios suggests a formation radius rather close to the star, below 5 stellar radii.\\

With its soft X-ray spectrum and its narrow X-ray lines, \tc\ clearly appears different from O-type stars. Comparing \tc\ to other B-type objects with high-resolution spectra, we find that the temperature distribution is quite typical of `normal' B stars: the DEM is similar to that of $\epsilon$\,Ori and only slightly broader than for $\beta$\,Cru and Spica \citep{zhe07}. Similar temperatures, fluxes and overall line properties were also found in a detailed analysis of $\beta$\,Cru \citep{coh08}, though in the latter case, more precise constraints could be found on the formation radius. On the other hand, \tc\ seems very different from $\tau$ Sco as far as softness, abundances, and flux level are concerned, but both objects present unshifted, narrow X-ray lines \citep{mew,coh03}. Such lines are at odds with the wind-shock model, though it is still unclear how this model could apply in the case of low mass-loss rates, as those of B-type stars. Actually, narrow lines are often attributed to magnetic confinement. However, this process is still uncertain for \tc, as magnetic field searches were inconclusive up to now \citep{hub08}. As for $\beta$\,Cru, additional observations, especially polarimetric ones, are requested before \tc\ can be fully understood and the origin of its X-ray emission pinpointed.

\begin{acknowledgements}
We acknowledge support from the Fonds National de la Recherche Scientifique (Belgium) and the PRODEX XMM and Integral contracts. We thank S. Hubrig for giving us the opportunity to read her paper before acceptance, and T. Morel for useful discussions.
\end{acknowledgements}

\end{document}